\documentclass[11pt,preprint]{aastex}
\usepackage{epsfig}
\usepackage{amsmath}
\usepackage{multirow}
\usepackage{color}

\begin{document}
\title{On the origin of $>10$ GeV photons in gamma-ray burst afterglows}
\author{Xiang-Yu Wang\altaffilmark{1,5}, Ruo-Yu Liu\altaffilmark{1,3,4,5}
and Martin Lemoine\altaffilmark{2}} \altaffiltext{1}{School of
Astronomy and Space Science, Nanjing University, Nanjing, 210093,
China}\altaffiltext{2}{Institut d'Astrophysique de paris, CNRS,
UPMC, 98 bis boulevard Arago, F-75014 Paris, France}
\altaffiltext{3}{Max-Planck-Institut f\"ur Kernphysik, 69117
Heidelberg, Germany } \altaffiltext{4}{Fellow of the International
Max Planck Research School for Astronomy and Cosmic Physics at the
University of Heidelberg (IMPRS-HD)}\altaffiltext{5}{Key
laboratory of Modern Astronomy and Astrophysics (Nanjing
University), Ministry of Education, Nanjing 210093, China}
\begin{abstract}
Fermi/LAT has detected long-lasting high-energy photons ($>100{\rm
MeV}$) from  gamma-ray bursts (GRBs), with the highest energy
photons reaching about 100 GeV. One proposed scenario is that they
are produced by high-energy electrons accelerated in  GRB forward
shocks via synchrotron radiation. We study the maximum synchrotron
photon energy in this scenario, considering the properties of the
microturbluence magnetic fields behind the shock, as revealed by
recent Particle-in-Cell simulations and theoretical analyses of
relativistic collisionless shocks. Due to the small-scale nature
of the micro-turbulent magnetic field, the Bohm acceleration
approximation, in which the scattering mean free path is equal to
the particle Larmor radius, breaks down at such high energies.
This effect leads to a typical maximum synchrotron photon of a few
GeV at 100 s after the burst and  this maximum synchrotron photon
energy decreases quickly with time. We show that the fast decrease
of the maximum synchrotron photon energy leads to a fast decay of
the synchrotron flux.
The 10-100 GeV photons detected after the
prompt phase can not be produced by the synchrotron mechanism.
They could originate from the synchrotron self-Compton emission of
the early afterglow if the circum-burst density is sufficiently
large, or from the external inverse-Compton process in the
presence of central X-ray emission, such as X-ray flares and
prompt high-latitude X-ray emission.

\end{abstract}

\section{Introduction}
The Fermi satellite has opened a new window at high energies in
studying gamma-ray bursts (GRBs). So far, Fermi large area
telescope (LAT) has detected high-energy photons  above 100 MeV
from more than 40 GRBs (Ackermann et al. 2013). The high-energy
emission ($>$100 MeV) generally lasts much longer than the prompt
KeV/MeV emission. The widely discussed scenario for this extended
GeV emission is the external shock model, where electrons are
accelerated by external forward shocks and produce GeV photons via
synchrotron radiation  (Kumar \& Barniol Duran 2009, 2010;
Ghisellini et al. 2009; Wang et al. 2010). The highest energy of
these photons reached about 100 GeV (in the local redshift frame)
during the early afterglows, as have been seen in GRB\,090902B,
GRB\,090926A,  and GRB 130427A (Abdo et al. 2009a;2009b; 2011a;
Zhu et al. 2013). This posed a challenge for the external shock
synchrotron scenario since the maximum synchrotron photon energy
for accelerated electrons under the most favorable condition (i.e.
the Bohm acceleration) is about 50 MeV in the shock rest
frame{\footnote{Kumar et al. (2012) suggest that particles can be
accelerated in the low background magnetic field region and
radiate in the high magnetic field region (i.e. in the amplified
turbulence magnetic field region), so the maximum synchrotron
emission can exceed 50 MeV. However, once the particles enter into
the low background magnetic field region, the particles can not be
scattered back to the upstream due to the lack of
micro-turbulence, and the acceleration will not continue. }} and
the bulk Lorentz factor of the external shock is $\Gamma\la300$ at
$\sim 100$ s after the burst trigger (Piran \& Nakar 2010; Barniol
Duran \& Kumar 2011; Sagi \& Nakar 2012). The maximum synchrotron
photon energy 50 MeV is obtained in the assumption of the fastest
acceleration -- the Bohm acceleration, where the acceleration time
is equal to the Larmor time of the particles in the magnetic field
(i.e. by equating the Larmor time with the synchrotron cooling
time in the magnetic field).

GRB afterglows are produced by relativistic collisionless shocks
expanding into a weakly magnetized ambient medium. In such shocks,
it is expected that the beam of super-thermal particles excite
micro-turbulence on plasma scales $c/\omega_{pi}$ in the ambient
medium (Medvedev \& Loeb 1999), where $\omega_{pi}=(4\pi n
e^2/m_p)^{1/2}$ is the plasma frequency. The Particle-in-Cell
(PIC) simulations indicate that the turbulence magnetic field
quickly relaxes behind the shock transition to { a
sub-equipartition} value of $\epsilon_B=0.01$ on spatial scales of
$\lambda\sim 10-30 c/\omega_{pi}$ (Chang et al. 2008; Keshet et
al. 2009). Such microturbulence is expected to decay (Gruzinov \&
Waxman 1999) on larger scales; these recent simulations
characterize the decay by a power law $\epsilon_B\propto
(l/\lambda_d)^{-\alpha_t}$ (where $l$ is the distance from the
shock front and {$\lambda_d$ is the decay
  length of the magnetic field downstream from the shock}). Modelling
of the broad band emission of LAT GRBs indeed points to the decay
of the magnetic field downstream of the shock, {with
  $-0.5\,\lesssim\,\alpha_t\,\lesssim -0.4$,} see Lemoine (2013) and
  Lemoine et al. (2013).

In relativistic collionless shocks responsible for the GRB
afterglows, the Bohm approximation may not be valid for maximum
energy electrons due to the small-scale nature of the
micro-turbulence generated in the shock layer (Kirk \& Reville
2009; Plotnikov et al. 2013; Lemoine 2013; Sironi et al. 2013),
i.e. the Larmor radius of the maximum energy electrons is much
larger than the length of the magnetic field, resulting a longer
time for electrons finishing a scattering cycle. The maximum
electron energy thus decreases, leading to a maximum synchrotron
photon energy significantly smaller than 50 MeV in the shock rest
frame and thus posing even more severe challenge for the
synchrotron scenario. The typical maximum synchrotron photon
energy, as shown in \S 2, is around GeV energies at hundreds of
seconds after the burst and this maximum energy decreases quickly
with time. We will show, in \S 3, that the decreasing maximum
photon energy could lead to a fast decay of the synchrotron
high-energy afterglow.

10-100 GeV photons detected after the prompt phase have energies
exceeding the maximum synchrotron photon energy, and thus must
have another origin. In \S 4, we study whether these highest
energy photons can be produced by the inverse Compton (IC)
emission in the afterglow, including the synchrotron self-Compton
emission from the afterglow and the external IC emission of
central X-ray emission. Finally we give our discussions and
conclusions in \S 5.

\section{The maximum  synchrotron photon energy}

The acceleration of particles in  GRB forward shocks is believed
to be governed by the Fermi mechanism. A non-thermal power-law
spectrum is formed, described by $dN/d\gamma_e\propto
\gamma_e^{-p}$, where $\gamma_e$ is the electron Lorentz factor.
Because the Lamor radius $R_L\gg \lambda$ for maximum energy
electrons, as shown below, particles only suffer a random small
angle deflection of the order of $\lambda/R_L$ as they cross a
coherence cell of size $\lambda$. Therefore, the Bohm
approximation breaks down and the residence time scale downstream
is given by $t_{res}=N t_{\lambda}=(R_L/\lambda)^2
(\lambda/c)=(R_L/\lambda)(R_L/c)$, where $N=(R_L/\lambda)^2$ is
the number of scatterings that the electrons experience before
returning to the upstream and $t_\lambda$ is the time spent in
crossing the coherence cell.

By equating the synchrotron cooling time $t_{syn}=6\pi m_\emph{e}
c/(\sigma_{\rm T}\gamma_e B^2)$ with the residence time $t_{res}$,
where $B$ is the magnetic field in the downstream region, we
obtain the Lorentz factor of the maximal energy electrons (Kirk \&
Reville 2009, Plotnikov et al. 2013; Sironi et al. 2013) , i.e.
\begin{equation}
\gamma_{e,max}=\left(\frac{6\pi \lambda e^2}{\sigma_{\rm T} m_e
c^2}\right)^{1/3}=2.5\times10^7 n_{-2}^{-1/6}\lambda_1^{1/3},\label{eq:gemax}
\end{equation}
where $\lambda\equiv 10 \lambda_1 c/\omega_{pi}$ and $n$ is the
number density of the surrounding interstellar medium. Note that
this Lorentz factor is independent of the magnetic field.

Since higher energy electrons travel a shorter distance before losing
their energy, they radiate their energy at the place closer to the
shock front. Thus, one expects that the highest energy synchrotron
photons are emitted by the highest energy electrons at the place { in
  very close proximity to} the shock front, while the low-frequency
emission with $\nu<\nu_c$, which are produced by electrons with
Lorentz factor $\gamma$ whose cooling timescale is longer than the
dynamic timescale, are emitted at the back of the shock. As the
microturbulence magnetic field decays with the distance from the
shock front, we assume $\epsilon_{B+}=0.01$ for the high-energy
emission detected by Fermi/LAT, while assume a value
$\epsilon_{B-}$ for low-energy afterglows, where $\epsilon_{B-}$
could be much smaller than $\epsilon_{B+}$  (see Lemoine 2013,
Lemoine et al. 2013 for details).

One can check that the Bohm approximation for $t_{res}$ fails at
this maximal Lorentz factor because
\begin{equation}
\frac{R_L(\gamma_{e,max})}{\lambda}=25 \lambda_1^{-2/3}
n_{-2}^{-1/24} \epsilon_{B+,-2}^{-1/2}E_{54}^{-1/8} t_2^{3/8} \gg
1.
\end{equation}
{Assuming a constant $\epsilon_{B+}$ parameter, one can derive the
time evolution of the maximum synchrotron photon energy from
Eq.~(\ref{eq:gemax}). For a constant external density, one derives
(see also Sironi et al. 2013)}

\begin{equation}
\begin{array}{ll}
\varepsilon_{\gamma, max}=\frac{1}{1+z}\frac{\Gamma
\gamma_{e,max}^2 e B}{2\pi m_e c}\\ =1.2 {\rm
GeV}(\frac{1}{1+z})^{1/4}
E_{54}^{1/4}n_{-2}^{-1/12}\epsilon_{B+,-2}^{1/2} \lambda_1^{2/3}
t_2^{-3/4}.
\end{array}
\end{equation}
{For the wind case, $\varepsilon_{\gamma, max}\propto t^{-2/3}$
(Sironi et al. 2013) . In this Letter, we assume that the
circumburst medium has a constant density, but the result in
section 3 holds also for the wind case. }

\section{Light curves of the $\ga 100$ MeV emission}
Since $\varepsilon_{\gamma, max}$ decreases quickly with time
($\propto t^{-3/4}$), the synchrotron flux in the LAT energy
window (i.e. $>100$ MeV) may decay faster than expected when
$\varepsilon_{\gamma, max}$ approaches to the low threshold energy
of the LAT window (i.e. 100 MeV). For a spectrum of $f_\nu \propto
\nu^{-\beta}$, the integrated flux from the LAT threshold energy
to the maximum energy is given by
\begin{equation}
F_{\rm LAT}\propto \left \{
\begin{array}{ll}
t^{-\alpha}(1-(\varepsilon_{\gamma,
max}/\varepsilon_0)^{-\beta+1}), \,\,\,\,\, \beta\ge 1\\
t^{-\alpha}{\rm ln}(\varepsilon_{\gamma, max}/\varepsilon_0),
\,\,\,\,\, \beta=1,
\end{array} \right.
\end{equation}
where $\alpha$ is the decay slope of the high-energy emission if
$\varepsilon_{\gamma, max}$ is infinity, which is typically
$\alpha=(3p-2)/4$  for high-energy gamma-ray emission.

We show in Fig.1 some examples of the light curve of the
synchrotron flux. We also calculate the self-synchrotron Compton
(SSC) component contribution to the flux above 100 MeV. As can be
seen from Fig.1, the relative strength between the synchrotron
component and the SSC component can lead to different shapes of
the observed light curves. In Fig. 1(a) and (b), the two
components add together to form a late flattening or an almost
straight power-law, even though the synchrotron component decays
quickly at a few tens of the deceleration time. When the density
of the circumburst medium is lower, the SSC component is not
important even at late times. In this case, we will see an
increasingly decaying synchrotron light curves,  as described by
Fig.1(c) and Fig. 1(d).

As an example, we fit the broad-band light curves of GRB\,080916C.
The LAT emission above 100 MeV decays quickly with a slope of
$\alpha=1.8$ {(Ackermann et al. 2013)}, which is much faster than
the expected from the standard synchrotron afterglow scenario. By
invoking the effect of the decreasing maximum photon energy, we
can reproduce such a fast decay, as shown in Fig.2. Note that the
high-energy emission before $\sim30$ s should be attributed to the
prompt emission. {A radiative blast wave interpretation for the
fast decay has also been proposed by Ghisellini et al. (2010).
This scenario needs a high value of the electron equipartition
factor (i.e., $\epsilon_e\simeq 1$) in addition to a fast-cooling
electron spectrum.}

\section{Inverse Compton origin for the 10-100 GeV photons }
As we have shown, the synchrotron photon energy is limited to be a
few GeV at $\sim 100$ seconds after the burst, so the 10-100 GeV
photons detected during the afterglow phase must have another
origin. Inverse-Compton  mechanism is then the most natural
mechanism. During the afterglow phase, there are two main sources
of the IC emission, one is the SSC emission of the forward shock
electrons and another is the external IC of central X-ray
emission, such as X-ray flares and  prompt high-latitude X-ray
emission, by the forward shock electrons. Here we first study
whether the SSC mechanism can explain these highest photons. Such
scenario has been discussed in Zhang \& M\'{e}sz\'{a}ros (2001)
for high-energy emission before the Fermi era.

\subsection{Synchrotron self-Compton mechanism }
In the SSC scenario, the highest energy photons are mainly
produced by  electrons with energy $\gamma<\gamma_c$ scattering
off afterglow photons, so we should use the magnetic field in the
back of the shock in the calculation (i.e.
$\epsilon_B=\epsilon_{B-}$). The value of $\epsilon_{B-}$ is found
to be $10^{-6}-10^{-4}$ from the broad-band fitting of the LAT GRB
data (e.g. Kumar \&  Barniol Duran 2010; Liu \& Wang 2011; He et
al. 2011; Lemoine et al. 2013). The cooling Lorentz factor and the
minimum Lorentz factor of electrons in forward shocks are given by
\begin{equation}
\gamma_c=6\times10^5\left(\frac{1+Y_c}{10}\right)^{-1}
E_{54}^{-3/8}n_{-2}^{-5/8}\epsilon_{B-,-4}^{-1}t_2^{1/8}(1+z)^{-1/8}
\\
\end{equation}
and
\begin{equation}
 \gamma_m=6\times10^3
f_p\epsilon_{e,-1}E_{54}^{1/8}n_{-2}^{-1/8}t_2^{-3/8}(1+z)^{3/8},\\
\end{equation}
respectively (Sari et al. 1998), where $Y_c$ is the Compton
parameter for electrons of energy $\gamma_c$ and $f_p\equiv
6(p-2)/(p-1)$. {As $\gamma_m<\gamma_c$ for typical parameter
values, the electron spectrum is in the slow-cooling regime.} The
SSC spectrum is characterized by two corresponding break
frequencies at
\begin{equation}
h\nu_m^{IC}=0.5 {\rm GeV} f_p^4\epsilon_{e,-1}^4
\epsilon_{B-,-4}^{1/2}E_{54}^{3/4}t_2^{-9/4}n_{-2}^{-1/4}(1+z)^{5/4}\\
\end{equation}
and
\begin{equation}
h\nu_c^{IC}=10^4 {\rm TeV}
\left(\frac{1+Y_c}{10}\right)^{-4}\epsilon_{B-,-4}^{-7/2}E_{54}^{-5/4}n_{-2}^{-9/4}t_2^{-1/4}(1+z)^{-3/4}.
\end{equation}
The Klein-Nishina scattering effect induces a break at
\begin{equation}
\begin{array}{ll}
h\nu_M=\Gamma\gamma_c m_e c^2\\=84 {\rm
TeV}\left(\frac{1+Y_c}{10}\right)^{-1}
E_{54}^{-1/4}n_{-2}^{-3/4}\epsilon_{B-,-4}^{-1}t_2^{-1/4}(1+z)^{1/4}.
\end{array}
\end{equation}

The flux at $h\nu_{\rm obs}=30 {\rm GeV}$ peaks when
$\nu_m^{IC}=\nu_{obs}$, which occurs at
\begin{equation}
t_p=20 {\rm s} f_p^{16/9}\epsilon_{e,-1}^{16/9}
\epsilon_{B-,-4}^{2/9}E_{54}^{1/3}n_{-2}^{-1/9}(\frac{1+z}{2})^{5/9}(\frac{h\nu_{\rm
obs}}{\rm 30 GeV})^{-4/9}\\
\end{equation}
This suggests that the SSC emission at 10-100 GeV peaks around
10-100 s for typical parameter values.

A convenient way to check whether the SSC emission can produce
enough $\ga10$ GeV photons is by comparing the number of SSC
photons to the synchrotron $\ga100$ MeV photons. The number of SSC
photons at $h\nu_1=30 {\rm GeV}$ is
$f_\nu^{IC}(\nu_1)=f_m^{IC}(\nu_1/\nu_m^{IC})^{-(p-1)/2}$ and the
number of synchrotron photons at $h\nu_2=100 {\rm MeV}$ is
$f_\nu^{syn}(\nu_2)=f_m^{syn}(\nu_c/\nu_m)^{-(p-1)/2}(\nu_2/\nu_c)^{-p/2}$,
where $f_m^{IC}$ and $f_m^{syn}$ are the peak flux of the
synchrotron component and SSC component, respectively. The two
peak fluxes are related by
\begin{equation}
\frac{f_m^{IC}}{f_m^{syn}}=\tau=\frac{1}{3}\sigma_{\rm T}
nR=2\times10^{-9} E_{54}^{1/4}n_{-2}^{3/4}(\frac{t_2}{1+z})^{1/4},
\end{equation}
where $\tau$ is the scattering optical depth by the shocked ISM,
$\sigma_{\rm T}$ is the Thomson scattering cross section and $R$
is the shock radius. Then, one can obtain
\begin{equation}
\frac{f_\nu^{IC}(\nu_1)}{f_\nu^{syn}(\nu_2)}=2\times10^{-4}\left(\frac{1+Y_c}{10}\right)(f_p\epsilon_{e,-1})^{p-1}
\epsilon_{B-,-4}^{3/4}E_{54}^{(3+p)/8}n_{-2}^{(11-p)/8}t_2^{(7-3p)/8}(1+z)^{3(p-1)/8}.
\end{equation}
Fermi/LAT detected one $\sim30$ GeV photon and  about 200 photons
above 100 MeV in both GRB\,090902B and GRB\,090926A, which implies
that $\frac{f_\nu^{IC}(\nu_1)}{f_\nu^{syn}(\nu_2)}\simeq
{\frac{1}{200}}$. Then one obtains
\begin{equation}
n\simeq0.3\left(\frac{1+Y_c}{10}\right)^{-\frac{8}{11-p}}(f_p\epsilon_{e,-1})^{-\frac{8(p-1)}{11-p}}
\epsilon_{B-,-4}^{-\frac{6}{11-p}}E_{54}^{-\frac{3+p}{11-p}}t_2^{-\frac{7-3p}{11-p}}(1+z)^{-\frac{3(p-1)}{11-p}}.
\end{equation}
For $\epsilon_{B-}$ in the range of $10^{-6}-10^{-4}$, the required
density is $n\simeq 0.3-10{\rm cm^{-3}}$. Therefore, only when the
density is sufficiently high, the SSC emission can explain the 10-100
GeV photons detected during the early afterglow. For GRB\,090902B,
however, modelling of the broadband (from radio to GeV) afterglows
gives a low value for the circumburst density with $n\sim 10^{-3}{\rm
  cm^{-3}}$ (Liu \& Wang 2011; Barniol Duran \& Kumar 2010; Cenko et
al. 2011; Lemoine et al. 2013), so the 30 GeV photon detected at
80 s after the trigger can not be interpreted as of SSC
origin~\footnote{ As discussed in Lemoine et al. (2013), a
  wind model does not provide as good a match to the data as a
  constant density medium for this burst, however it leads to a
  relatively high external wind parameter, for which SSC emission is
  expected to contribute significantly to the flux at $\gtrsim\,1$GeV;
  in this case, it might be possible to accommodate the origin of this
  highest energy photon with the SSC model.}. For GRB\,090926A,
because there is no radio flux constraint on the parameters, a high
density is allowed by the broadband data. We perform a broad-band
fitting of GRB\,090926A, including one 20 GeV photon at 25 s after the
burst, as shown in Fig.3. The SSC component at energies above 100 MeV
becomes dominant after 100 s after the trigger. At energies above 10
GeV, the SSC component is dominant and peaks at 30-50 s after the
trigger, while the synchrotron flux is negligible. Note that the
early-time high-energy emission before $\sim20$ s should be attributed
to the prompt emission.

GRB130427A shows high-energy photons above 100 MeV up to about one
day and high-energy photons above 30 GeV after several thousands
seconds after the burst (Zhu et al. 2013). The situation is
similar to GRB 940217, in which a 18 GeV photon was detected by
Energetic Gamma-Ray Experiment Telescope (EGRET), at about 5000
seconds after the burst (Hurley et al. 1994). This poses severe
challenge for the synchrotron afterglow model.  As such late
times, the SSC mechanism is the likely mechanism for $\ga$ 10 GeV
photons,  corresponding to the case in Fig.1(a).

\subsection{External IC of central X-ray emission}
During the early afterglow, X-ray emission from the central source
has been usually observed by Swift. During the first hundreds of
seconds, steeply decaying X-ray emission from the high-latitude
region of the emitting shells are common (Tagliaferri et al.
2005). X-ray flares are also common during the early afterglow,
with a fraction of $\ga30\%$ (Nousek et al. 2006; Chincarini et
al. 2010). We study whether the external IC of the central X-ray
emission can explain the 90 GeV photon seen in e.g. GRB\,090902B
if the surrounding circumburst density is low. We use X-ray flare
emission as an illustration.

Assuming that an X-ray flare occurs at $t=100t_2$ s after the
burst with a duration of $\delta t/t\sim 0.3$ and the peak of the
X-ray flare is $\varepsilon_{\rm pk}=1{\rm keV}$, the peak  of the
observed EIC $\nu f_{\nu}$ flux is
\begin{equation}
h\nu_m^{\rm EIC}=2\gamma_m^2\varepsilon_{\rm pk}= 70{\rm GeV}
f_p^2\epsilon_{e,-1}^2
E_{54}^{1/4}n_{-3}^{-1/4}t_2^{-3/4}(\frac{1+z}{2})^{-1/4}.
\end{equation}
So this peak energy can reach up to 100 GeV at  $\la100$ s after
the trigger for typical parameters of LAT GRBs. The X-ray flare
photons cause enhanced cooling of electrons through the IC
process. According to Wang et al. (2006), when the X-ray flare
flux is larger than a critical flux
\begin{equation}
F_{X,c}=3\times10^{-7}E_{54}^{1/2}f_p^{-1}\epsilon_{e,-1}^{-1}n_{-3}^{-1/2}t_2^{-1/2}D_{28}^{-2}{\rm
erg cm^{-2} s^{-1}},
\end{equation}
we have $\gamma_m\ga\gamma_c$ (for $p=2.3$),  and thus all the
newly shocked electrons will cool, emitting most of their energy
into the IC emission. The peak spectral flux of the EIC emission
is
\begin{equation}
\begin{array}{ll}
f_{\rm c}^{\rm EIC}=\tau\left(\frac{F_{\rm X}}{\varepsilon_{\rm
pk}}\right) k_{a}\\ =1.2\times10^{-34}k_{a}
E_{54}^{1/4}n_{-3}^{3/4}t_{2}^{1/4}F_{\rm X,-7} \,{\rm
erg\,cm^{-2}\,s^{-1}\,Hz^{-1}},
\end{array}
\end{equation}
where $k_{a}\sim0.4$ is the correction factor accounting for the
suppression of the IC flux due to the anisotropic scattering
effect compared to the isotropic scattering case (Fan \& Piran
2006; He et al. 2009). Therefore, the  photon flux of the EIC
emission at $h\nu_m^{\rm EIC}$ for the fast cooling case (i.e.,
$\nu_{\rm c}^{\rm EIC}<\nu_{\rm m}^{\rm EIC})$ is ( He et al.
2012)
\begin{equation}
n_{\rm m}^{\rm EIC} =10^{-6}k_a f_p^{-1}\epsilon_{e,-1}^{-1}
E_{54}^{3/4}n_{-3}^{1/4}t_2^{-1/4}D_{L,28}^{-2}{\rm ph
\,cm^{-2}\,s^{-1}}.
\end{equation}
We find that the  photon flux is insensitive to the density of the
circumburst medium, in stark contrast to the SSC mechanism. With
the LAT effective area of $A=10^4{\rm cm^2}$ for  $>10$ GeV
photons, the number of  high-energy photons of $h\nu_m^{\rm EIC}$
detected by LAT { during the flare period $\delta t=0.3 t$} is
estimated to be
\begin{equation}
N=n_{\rm m}^{\rm EIC} A \delta t \simeq 0.1
f_p^{-1}\epsilon_{e,-1}^{-1}
E_{54}^{3/4}n_{-3}^{1/4}D_{L,28}^{-2}\, {\rm photons}.
\end{equation}
Thus, for strong GRBs with $E\ga 10^{55}{\rm erg}$ , the external
IC emission could produce  10-100 GeV photons in the presence of
strong central X-ray emission, even the density of the circumburst
medium is low. This could explain the 90 GeV (in the local
redshift frame) photon detected in GRB\,090902B at 80 s after the
trigger.

\section{Discussions and Conclusions}
We have shown that, considering the properties of the
microturbulence magnetic field in the relativistic GRB afterglow
shocks, the maximum synchrotron photon energy produced by
shock-accelerated electrons is  a few GeV at hundreds of seconds
after the burst and this maximum photon energy decreases quickly
with time. Thus, the synchrotron afterglow scenario can only
explain the extended high-energy emission below several GeV. The
fast decrease of the maximum photon energy could lead to a fast
decay of the synchrotron light curves, as seen in some Fermi/LAT
GRBs.

High energy photons above 10 GeV, seen in some strong GRBs, should
have another origin. Inverse-Compton mechanism is the natural
mechanism for such photons. The afterglow SSC mechanism is one
possible scenario if the circumburst medium density is
sufficiently large (i.e. typically $\ga0.3{\rm cm^{-3}}$), which
may explain $>10$ GeV photons in a few cases, such as GRB\,090926A
and GRB130427A. An alternative mechanism is the external IC of
central X-ray emission, such as X-ray flares or prompt
high-latitude X-ray emission, since the IC flux from this process
is insensitive to the density of the circumburst medium and that
central X-ray emission is common during the early afterglows.
Interestingly, this scenario has been supported by the
simultaneous detections of X-ray flares (by Swift) and GeV
emission (by Fermi LAT) in GRB\,100728A (Abdo et al. 2011b; He et
al. 2012).

High energy photons above 10 GeV have also been detected from GRBs
during the prompt phase, such as in GRB080916C and GRB090510. As
the high-energy emission lies at the extrapolation of the Band
high-energy component, e.g. in GRB080916C (Abbo et al. 2009c), it
was suggested that these high energy photons could be produced by
synchrotron radiation in the Bohm regime (Wang et al. 2009). Since
the internal shock/dissipation mechanism that produces the prompt
emission {involves mildly relativistic shocks in a possibly
magnetized environment, the physics of acceleration is expected to
be different, and } the Bohm acceleration may be possible for the
highest electrons producing the prompt $\ga10$ GeV photons.

\acknowledgments This work is supported by the 973 program under
grant 2009CB824800, the NSFC under grants 11273016,  10973008, and
11033002, the Excellent Youth Foundation of Jiangsu Province
(BK2012011), the Fok Ying Tung Education Foundation { and the
  PEPS/PTI program of the INP (CNRS)}.

\clearpage
\begin{figure}
\plotone{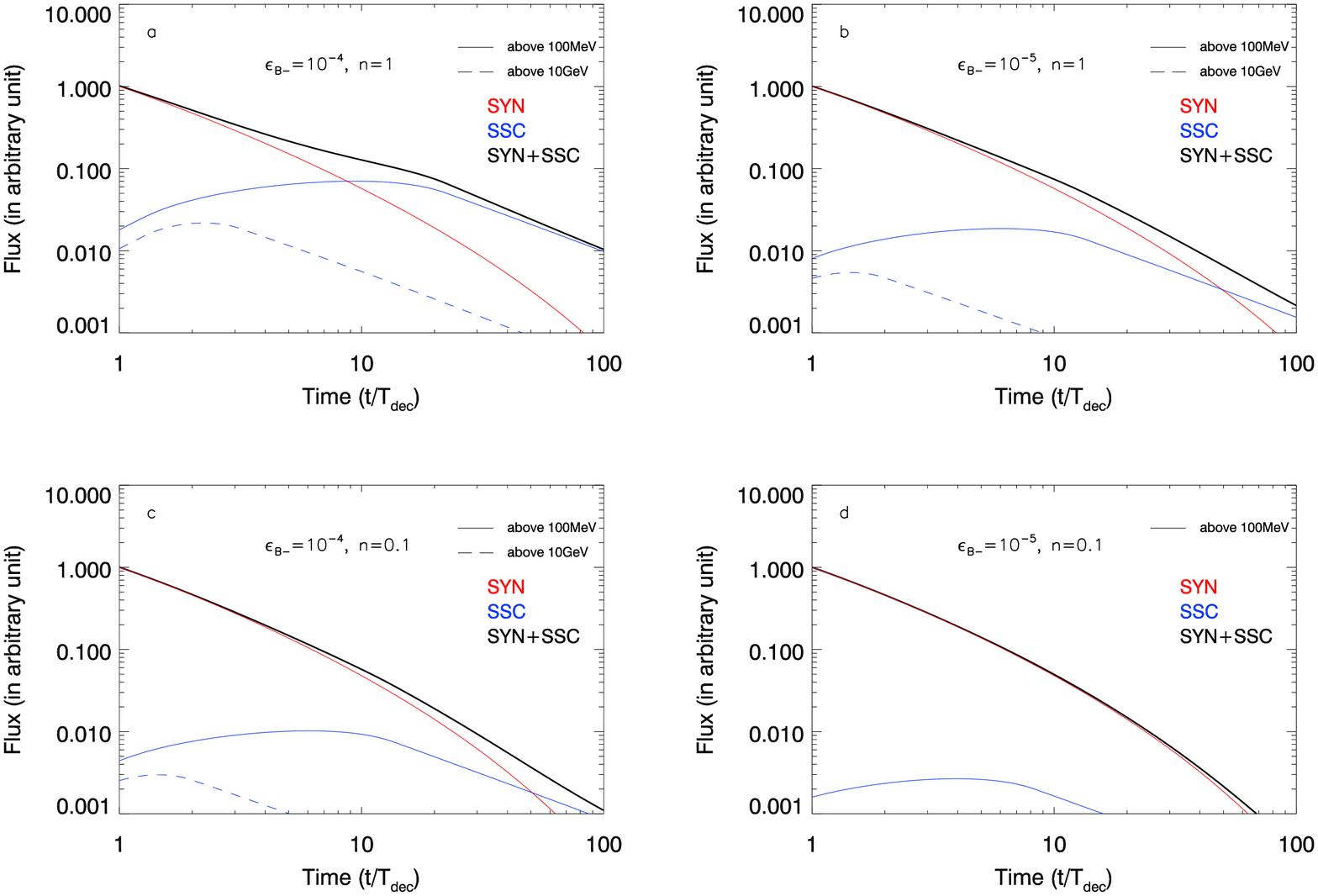} \caption{Predicted light curves of the
high-energy emission above 100 MeV (the solid lines) and above 10
GeV (the dashed lines). The total flux above $100{\rm MeV}$ (the
black solid lines) is the sum of the flux of the afterglow
synchrotron component (the red solid lines) and the synchrotron
self-Compton component (the blue solid lines). Note that the
synchrotron flux above 10 GeV is too low to be visible in the
figure. The parameter values are chosen as $E=10^{55}{\rm erg}$,
$\epsilon_e=0.1$, $\epsilon_{B+}=0.01$, $p=2.2$. The parameter
values of $\epsilon_{B-}$ and $n$ are shown in the figure.}
\end{figure}

\clearpage
\begin{figure}
\plotone{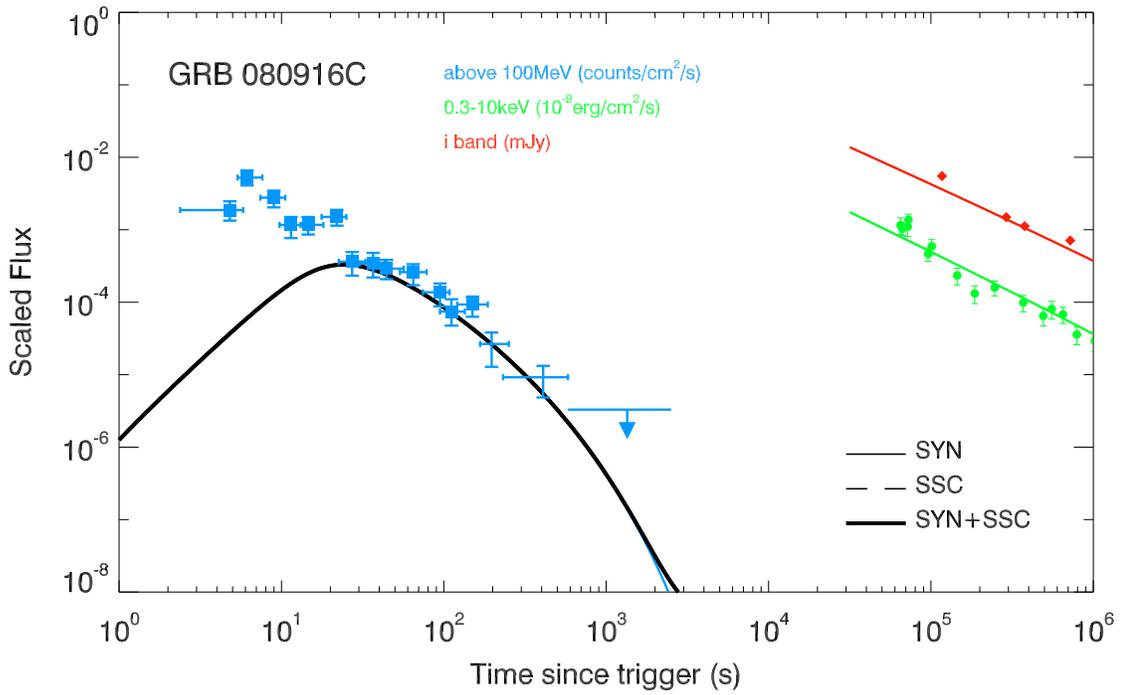} \caption{Fit of the broadband afterglow data of
GRB080916C. The solid lines represent the synchrotron flux at
high-energy ($>100$ MeV) , X-ray and optical frequencies. The SSC
flux at $>100$ MeV is too low to be visible. The parameter values
are $E=10^{55}{\rm erg}$, $n=0.003{\rm cm^{-3}}$,
$\epsilon_e=0.5$, $\epsilon_{B+}=0.01$,
$\epsilon_{B-}=3\times10^{-6}$, $\Gamma_0=800$ and $p=2.3$. }
\end{figure}
\begin{figure}
\plotone{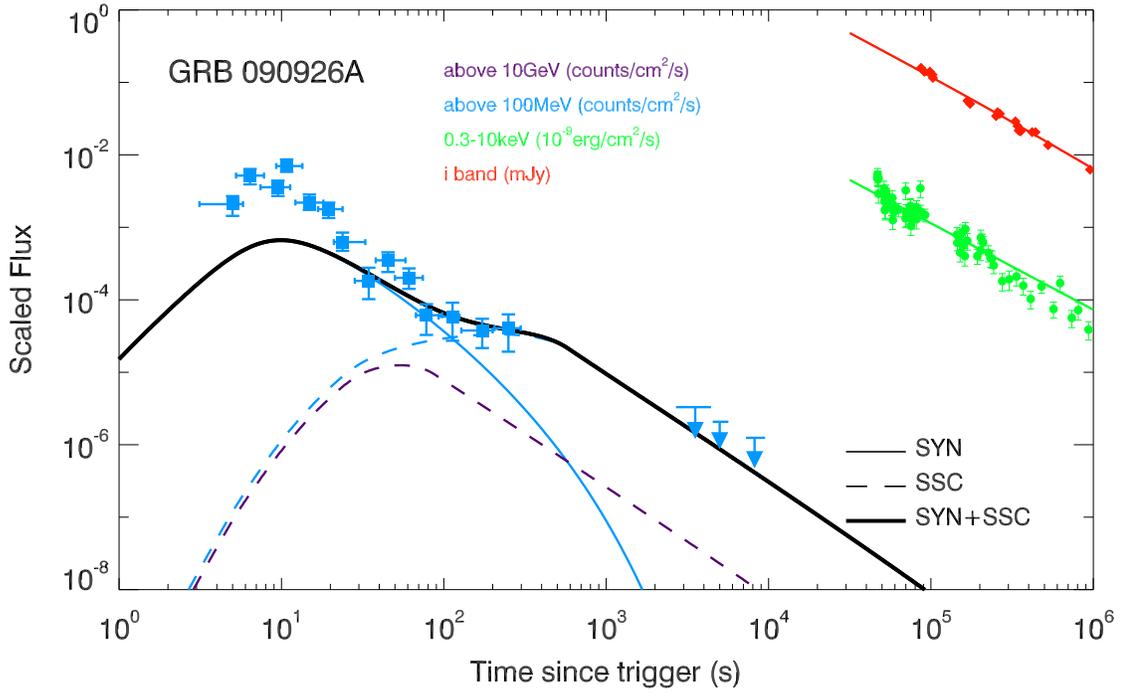} \caption{Fit of the broadband afterglow data of
GRB090926A. The blue solid line represents the synchrotron flux
and the blue dashed line represents the SSC flux above 100 MeV,
while the black solid line represents the sum of them. The purple
dashed line represents the SSC flux above 10 GeV. The parameter
values are $E=2\times10^{55}{\rm erg}$, $n=1.2{\rm cm^{-3}}$,
$\epsilon_e=0.1$, $\epsilon_{B+}=0.005$,
$\epsilon_{B-}=6\times10^{-6}$, $\Gamma_0=600$ and $p=2.5$. }
\end{figure}

\end{document}